\documentclass[
    ,final            
  ,numberedheadings 
  ]
  {aipproc}
\usepackage{astro_bib_macro}
\layoutstyle{8x11double}

\newcommand{\nebu}{}\def\nebu/{NEBU\_3D}
\newcommand{\ion}[2]{#1~\textsc{#2}}
\newcommand{\mysp}{}\def\mysp/{}
\newcommand{\alloa}[3]{\ion{#1\mysp/}{#2}\ #3\AA}
\newcommand{\forba}[3]{[\ion{#1\mysp/}{#2}]\ #3\AA}

\newcommand{\dforba}[4]{[\ion{#1\mysp/}{#2}]\ #3,#4\AA}

\newcommand{\hbeta}{}\def\hbeta/{H$\beta$}

\begin{document}

\title{NEBU\_3D: A fast pseudo-3D photoionization code for aspherical planetary nebulae and HII regions}


\author{C. Morisset}{
  address={Instituto de Astronom\'{\i}a, Universidad Nacional
           Aut\'onoma de M\'exico; Apdo. postal 70--264; Ciudad Universitaria;
           M\'exico D.F. 04510; M\'exico. Morisset@astroscu.UNAM.mx}
}

\author{G. Stasi\'nska}{
  address={LUTH, Observatoire de Meudon, 5 place J. Janssen,
           F-92195 Meudon Cedex, France.}
}

\author{M. Pe\~na}{
  address={Instituto de Astronom\'{\i}a, UNAM}
}

\begin{abstract}
We describe a  pseudo-3D photoionization code, \nebu/ and its associated visualization tool, VIS\_NEB3D, which are able to easily and rapidly treat a wide variety of nebular geometries, by combining models obtained with a 1D photoionization code. We also present a tool, VELNEB\_3D, which can be applied to the results of 1D or 3D photoionization codes to generate emission line profiles, position-velocity maps and 3D maps in any emission line by assuming an arbitrary velocity field. 
As examples of the capabilities of these new tools, we consider three very different theoretical cases. The first one is a blister HII region, for which we have also constructed a spherical model (the spherical impostor) which has exactly the same H$\beta$ surface brightness distribution as the blister model and the same ionizing star. The second example shows how complex line profiles can be obtained even with a simple expansion law if the nebula is bipolar and the slit slightly off-center.
The third example shows different ways to produce line profiles that could be attributed to a turbulent velocity field while there is no turbulence in the model.
\end{abstract}
\maketitle

\section{The need for 3D photoionization codes}

Most ionized nebulae have complex morphological structures. This is true not only for HII regions (Kim \& Koo 2003, {{Reich} et~al.} {1990}, {{O'Dell}} {2001},{{Mart{\'{\i}}n-Hern{\' a}ndez} et~al.} {2003}), but also for planetary nebulae (see many contributions to this book). High spatial resolution images and since a few years Integral Field Units observations are giving more information about the morphologies and kinematics of HII regions and PNe.

To model such aspherical structures, a 3D photoionization codes is needed. A few of them now exist ({{Baessgen} et~al.} {1995}, {{Gruenwald} et~al.} {1997}, {{Ercolano} et~al.} {2003}, {{Wood} et~al.} {2004}).
But there are some practical difficulties with 3D model fitting of real nebulae. One is the enormous parameter space to consider, especially in the case of complex morphologies. Another one is the fact that 3D photoionization models are very CPU-time and memory consuming: several hours are needed on clusters or supercomputers to make a model (even if the time and memory can be reduced when modeling simpler geometries). 

Also, the difficulty in visualising the results does not help.
\section{The tools NEBU\_3D, VISNEB\_3D and VELNEB\_3D}

In this paper, we present a pseudo-3D photoionization code (NEBU\_3D), which models aspherical planetary nebulae in a few minutes. 
The code uses various runs of a 1D photoionization code (NEBU, see {{Morisset} \& {P\'equignot}} {1996}, {{P{\' e}quignot} et~al.} {2001}), changing the nebular density distribution at each run according to an angular variation law. 
The set of the 1D outputs (such as the electron temperature, the emissivities of selected lines at each radial position, etc...), which is defined according to the goal of the model, is then read and interpolated on a predefined coordinate cube.

Its associated visualisation tool, VISNEB\_3D, allows one to easily plot diagrams that are useful for understanding the physical conditions inside the models and diagrams that can be directly compared to observations. The resulting emissivity cubes (one for each line of interest) can then be rotated and a projection on the sky plane is done to obtain surface brightness maps.

With such a procedure, one can  easily construct models of ellipsoidal or bipolar nebulae. One can also treat champagne flows or even irregular nebular geometries provided that the photoionization source is unique and that the effects of non-radial ionization are not expected to be important.

We also present a tool, VELNEB\_3D, which can be applied to the results of 3D photoionization codes to generate emission line profiles, position-velocity maps and 3D maps in any emission line by assuming an arbitrary velocity field. More details can be found in {{Morisset} \& {Stasi{\' n}ska}} (2005).

\nebu/ is thus not a full 3D photoionization code like Mocassin ({{Ercolano} et~al.} {2003}). The main limitation of \nebu/ is in the treatment of the diffuse field, which is not fully consistent. For example, \nebu/ should not be used to model the ionization of shadows behind knots, as 1D models cannot account in an accurate way for the ionization of the shadowed region by radiation produced by the surrounding material.
On the other hand, when the radiation field is not changing drastically from one direction to the other, the use of a set of 1D models does not lead to strong errors. The main advantage of \nebu/ is that it is faster by 3 to 4 orders of magnitude than a full 3D code when running on the same computer.

More details can be found in Morisset et al. (2005) and Morisset \&  Stasi{\' n}ska (2005).

\section{Applications of NEBU\_3D: a few toy models}

 In the following subsections a few examples of models obtained by \nebu/ are discussed to illustrate the capacities of 3D modelling.

\subsection{A blister and its spherical impostor}

A blister model is obtained assuming a plan parallel density distribution. In the case of a face-on blister, the surface brightness map shows a decrease of the emission  with the angular distance to the star; the same can be obtained with a spherical distribution around the ionising star. Such a distribution is called the spherical impostor.

The blister model presented here is a plane parallel slab, with density increasing exponentially with the distance to the ionizing star, as found by {{Wen} \&
  {O'Dell}} {1995} for the Orion nebula. The gas density distribution law of the spherical impostor has been obtained by trial and error, so that the HI surface brightness is the same than for the blister seen pole-on.

\begin{figure}
  \includegraphics[height=.22\textheight]{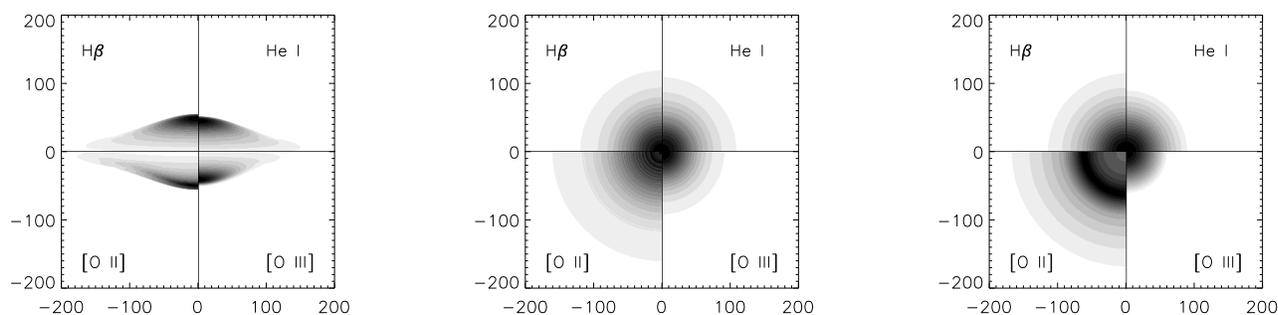}
  \caption{Synthetic H$\beta$, \alloa{He}{i}{6678}, \dforba{O}{ii}{3726}{29} and \forba{O}{iii}{5007} surface brightness maps from upper left to lower right. Left panel: the blister model seen tangentially (in this case the maps for \dforba{O}{ii}{3726}{29} and \forba{O}{iii}{5007} are upside down); central panel: the blister model seen face on; right panel: the spherical impostor model.}\label{fig:blister.images}
\end{figure}

The Fig.~\ref{fig:blister.images} shows surface brightness maps for the blister seen face-on (left panel), seen pole-on (center panel) and for the spherical impostor (right panel), for four different lines. The difference between the blister and its impostor is perceptible for the low ionization species, where a ring is visible in the case of the impostor. Emission line profiles are also different, see Morisset et al. (2005).

\subsection{A bipolar nebula: complexe line profiles}

\begin{figure}
  \includegraphics[height=.25\textheight]{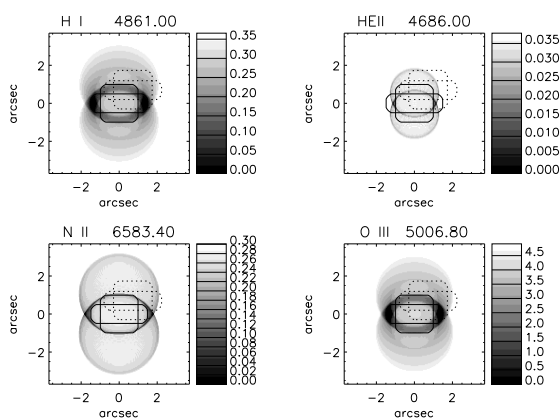}
  \caption{Monochromatic images obtained for three emission lines (\hbeta/, \alloa{He}{ii}{4689}, \forba{N}{ii}{6583}, and \forba{O}{iii}{5007}), for the bipolar nebula. The apertures used to compute the emission line profiles shown in Fig.~\ref{fig:bipol.profiles} are superimposed to the images: the sizes of the apertures are: 1''x3'' and 2''x2''. The largest aperture (10''x10'') covers the entire nebula and is not shown. The centered apertures are drawn with solid lines. Off-center apertures  are drawn with dashed lines. Surface brightness units are arbitrary, but the same for the four images.}\label{fig:bipol.apertures}
\end{figure}

\begin{figure}
  \includegraphics[height=.34\textheight]{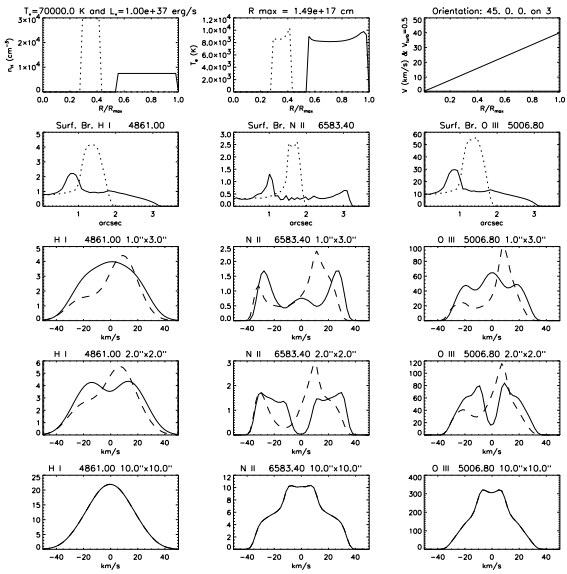}
  \caption{Some physical parameters, surface brightness distribution and line profiles for the bipolar nebula. Upper row:
radial distribution of the hydrogen density (left panel), the electron temperature (middle panel) and the expansion velocity (right panel). Solid curves correspond to the polar direction, dotted curves to a perpendicular one. Second row : from left to right, surface brightness distribution in the  \hbeta/, \forba{N}{ii}{6583} and \forba{O}{iii}{5007} lines along the polar axis (solid curves) and in the perpendicular direction (dotted curves). 
Last three rows: from left to right, line profiles \hbeta/, \forba{N}{ii}{6583} and \forba{O}{iii}{5007} lines, through apertures represented in Fig.~\ref{fig:bipol.apertures} (the size of the aperture is specified on top of each plot). Solid curves correspond to centered apertures, dashed curves to off-centered apertures. Intensity units are arbitrary, but the same for all the plots.}\label{fig:bipol.profiles}
\end{figure}

A bipolar nebula is constructed and a linear expansion velocity law is assumed. Monochromatic images are shown for four lines in Fig.~\ref{fig:bipol.apertures}. Emission line profiles are obtained through various apertures and are plotted in Fig.~\ref{fig:bipol.profiles}. We can see that complex profiles (especially in case of narrow off-center apertures) are obtained using a very simple expansion law, the complexity issuing from the bipolar morphology. 

\begin{figure}
  \includegraphics[height=.25\textheight]{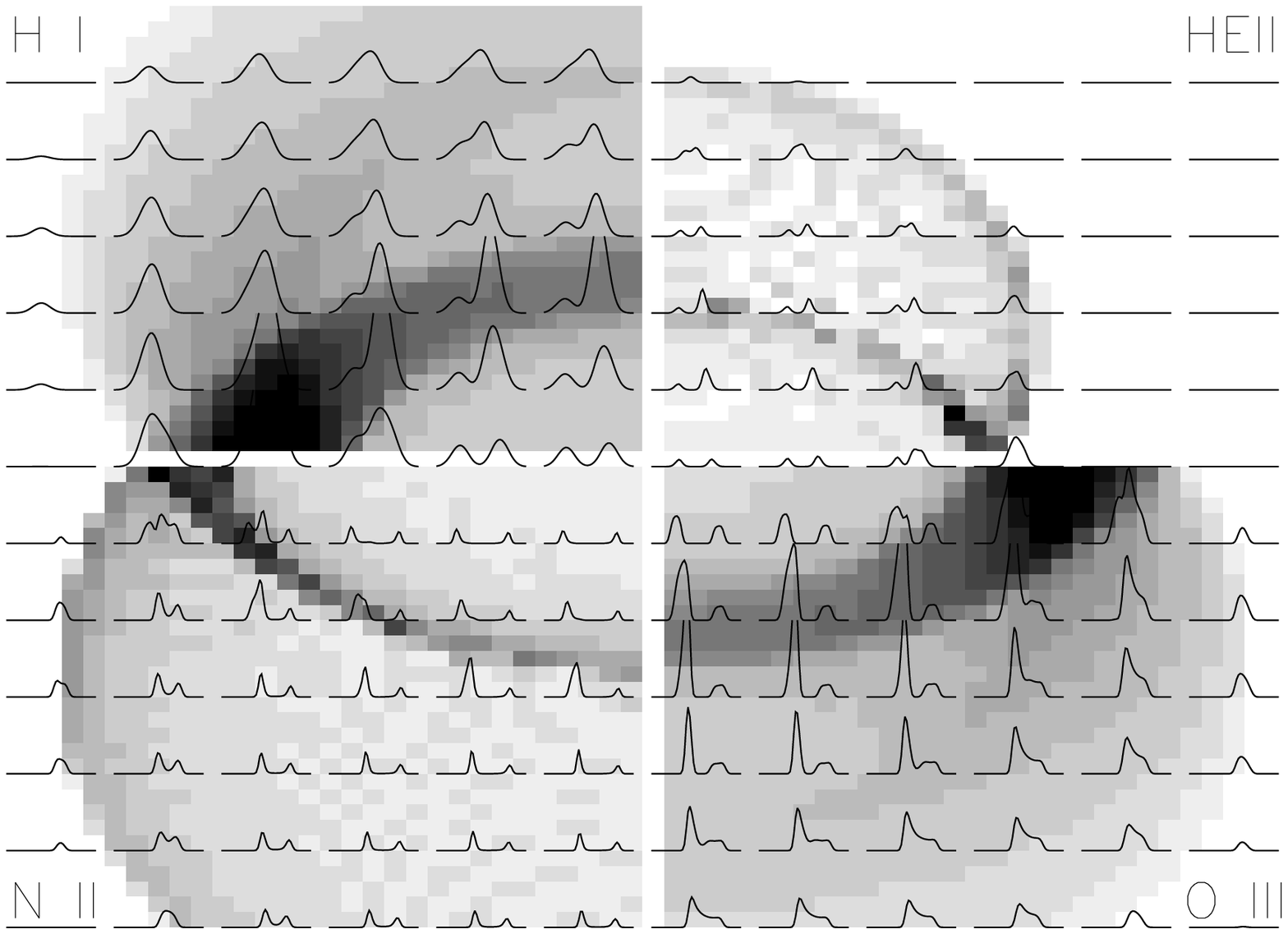}
  \caption{A 3D map of the bipolar nebula model. Each quadrant corresponds to a different emission line. The surface brightness is represented by levels of grey. Superimposed on this image are the profiles of the same lines, integrated over areas of 5x5 pixels.}\label{fig:bipol.IFU}
\end{figure}

There are now an increasing number of instruments which use integral field units (IFU) to achieve spectroscopy of an extended portion of the sky, either using lens arrays, optical fibres or image slicers. These instruments can thus generate 3D maps of extended objects, with wavelength (or velocity) being the third dimension. Such examples can be found in {{Ambrocio-Cruz} et~al.} (2004) and {{Vasconcelos} et~al.} (2005). Our tool is well-suited to also produce such maps for photoionization models of asymmetric nebulae. One such example is shown in Fig.~\ref{fig:bipol.IFU}, which represents the bipolar nebula model.

\subsection{An ellipsoidal nebula mimicking turbulence}

Recently, {{Neiner} et~al.} (2000) and {{Gesicki} et~al.} (2003) have claimed to find evidence for turbulence in certain planetary nebulae, especially in those ionized by [WC] type central stars. 

Two models were constructed, the first one being a spherical nebula with linear expansion law and high turbulence, the second one being an ellipsoidal nebula seen pole-on with a linear expansion law and no turbulence. With such a configuration, the weight of the zones of nearly zero radial velocity becomes large, and high velocity wings are produced as well. 

\begin{figure}
  \includegraphics[height=.34\textheight]{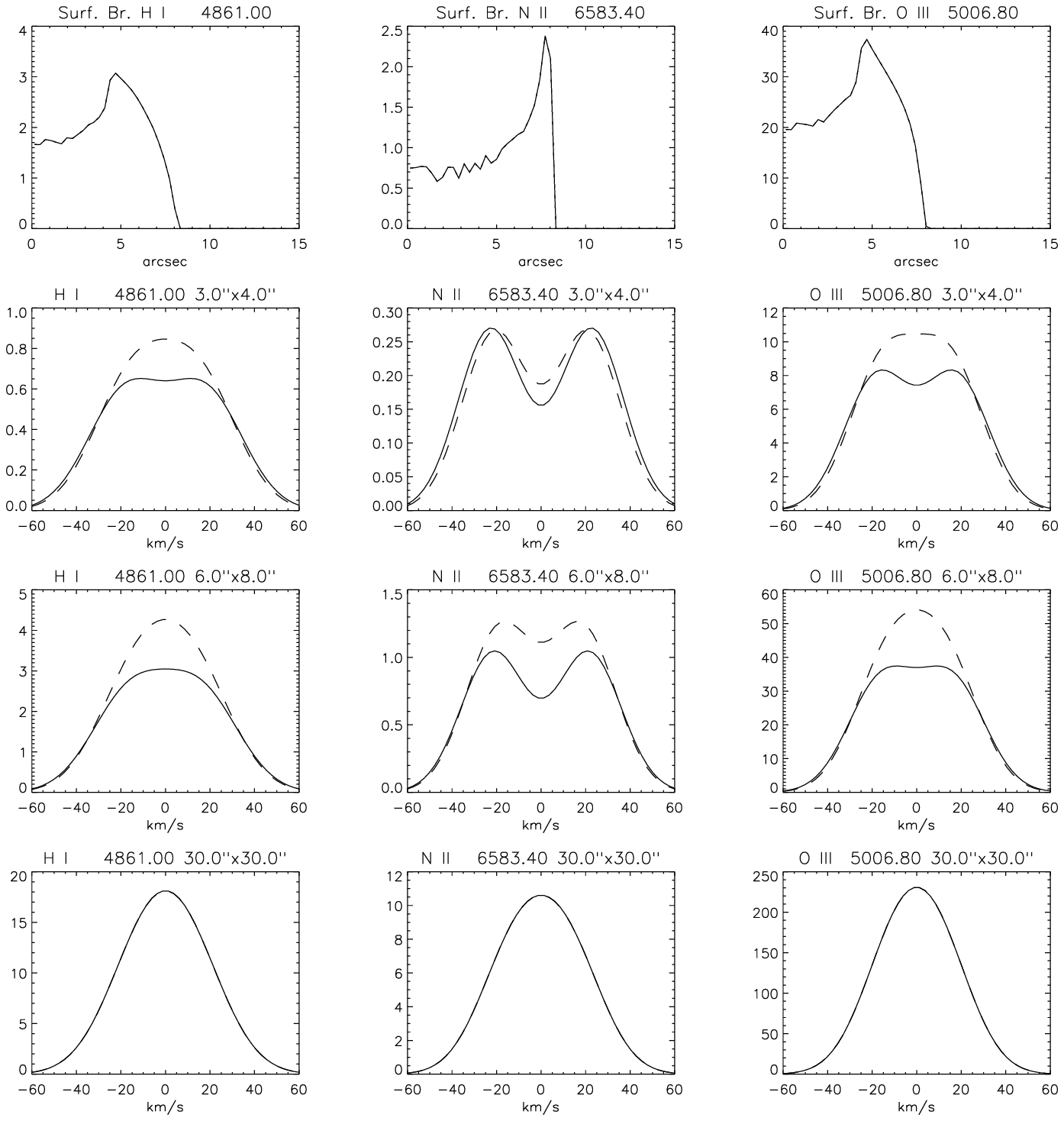}
  \caption{Spherical nebula, with a low velocity Hubble flow expansion law and high turbulence. The setup of the figure is the same as for Fig.~\ref{fig:bipol.profiles}}\label{fig:sphere.profiles}
\end{figure}

\begin{figure}
  \includegraphics[height=.34\textheight]{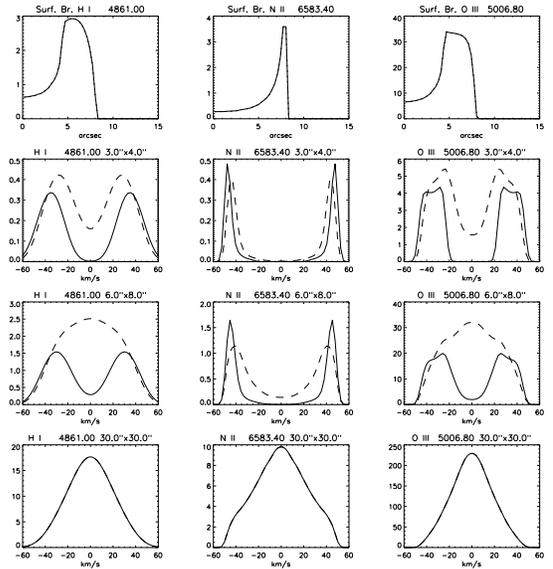}
  \caption{Ellipsoidal nebula, seen pole-on, with a Hubble flow expansion law and without turbulence. The setup of the figure is the same as for Fig.~\ref{fig:bipol.profiles}}\label{fig:ellips.profiles}
\end{figure}

Therefore, nebulae that would correspond to such models would be impossible to distinguish in practise, if using observational setups such as those represented by the last two rows of Figs.~\ref{fig:sphere.profiles} and ~\ref{fig:ellips.profiles}. Note, however, that with a smaller slit, the two models can be distinguished, at least if the slit is perfectly centered (compare the second rows of Figs.~\ref{fig:sphere.profiles} and ~\ref{fig:ellips.profiles}) because in that case the splitting of the emission lines due to crossing two emitting regions on the line of sight. 

\section{Future applications}

\nebu/ can easily be adapted to any other 1D photoionization code, a version using Cloudy ({{Ferland} et~al.} {1998}) is in development and a Cloudy\_3D will be public within a few months.

\begin{theacknowledgments}
G.S. is grateful to the Instituto de Astronomia, UNAM, Mexico, for hospitality and financial support. The computations were carried out on a AMD-64bit computer financed by grant PAPIIT IX125304 from DGAPA (UNAM,Mexico). C.M. is partly supported by grant Conacyt-40095 (Mexico).
\end{theacknowledgments}


\begin{thebibliography}{}

\bibitem[\protect\citeauthoryear{{Ambrocio-Cruz}, {Laval}, {Rosado},
  {Georgelin}, {Marcelin}, {Comeron}, {Delmotte} \& {Viale}}{{Ambrocio-Cruz}
  et~al.}{2004}]{2004AJ....127.2145A}
{Ambrocio-Cruz} P.,  {Laval} A.,  {Rosado} M.,  {Georgelin} Y.~P.,  {Marcelin}
  M.,  {Comeron} F.,  {Delmotte} N.,    {Viale} A.,  2004, \aj, 127, 2145

\bibitem[\protect\citeauthoryear{{Baessgen}, {Diesch} \& {Grewing}}{{Baessgen}
  et~al.}{1995}]{BDG95}
{Baessgen} M.,  {Diesch} C.,    {Grewing} M.,  1995, \aap, 297, 828

\bibitem[\protect\citeauthoryear{{Ercolano}, {Barlow}, {Storey} \&
  {Liu}}{{Ercolano} et~al.}{2003}]{Erc03}
{Ercolano} B.,  {Barlow} M.~J.,  {Storey} P.~J.,    {Liu} X.-W.,  2003, \mnras,
  340, 1136

\bibitem[\protect\citeauthoryear{{Ferland}, {Korista}, {Verner}, {Ferguson},
  {Kingdon} \& {Verner}}{{Ferland} et~al.}{1998}]{1998PASP..110..761F}
{Ferland} G.~J.,  {Korista} K.~T.,  {Verner} D.~A.,  {Ferguson} J.~W.,
  {Kingdon} J.~B.,    {Verner} E.~M.,  1998, \pasp, 110, 761

\bibitem[\protect\citeauthoryear{{G{\' o}rny}, {Schwarz}, {Corradi} \& {Van
  Winckel}}{{G{\' o}rny} et~al.}{1999}]{GSCW99}
{G{\' o}rny} S.~K.,  {Schwarz} H.~E.,  {Corradi} R.~L.~M.,    {Van Winckel} H.,
   1999, \aaps, 136, 145

\bibitem[\protect\citeauthoryear{{Gesicki}, {Acker} \& {Zijlstra}}{{Gesicki}
  et~al.}{2003}]{GAZ03}
{Gesicki} K.,  {Acker} A.,    {Zijlstra} A.~A.,  2003, \aap, 400, 957

\bibitem[\protect\citeauthoryear{{Gruenwald}, {Viegas} \&
  {Broguiere}}{{Gruenwald} et~al.}{1997}]{GVB97}
{Gruenwald} R.,  {Viegas} S.~M.,    {Broguiere} D.,  1997, \apj, 480, 283

\bibitem[\protect\citeauthoryear{{Kim} \& {Koo}}{{Kim} \& {Koo}}{2003}]{Kim03}
{Kim} K.,  {Koo} B.,  2003, \apj, 596, 362

\bibitem[\protect\citeauthoryear{{Manchado}, {Guerrero}, {Stanghellini} \&
  {Serra-Ricart}}{{Manchado} et~al.}{1996}]{MGSS96}
{Manchado} A.,  {Guerrero} M.~A.,  {Stanghellini} L.,    {Serra-Ricart} M.,
  1996, Bulletin of the American Astronomical Society, 29, 732

\bibitem[\protect\citeauthoryear{{Mart{\'{\i}}n-Hern{\' a}ndez}, {van der
  Hulst} \& {Tielens}}{{Mart{\'{\i}}n-Hern{\' a}ndez} et~al.}{2003}]{MHT03}
{Mart{\'{\i}}n-Hern{\' a}ndez} N.~L.,  {van der Hulst} J.~M.,    {Tielens}
  A.~G.~G.~M.,  2003, \aap, 407, 957

\bibitem[\protect\citeauthoryear{{Meixner}, {Kastner}, {Balick} \&
  {Soker}}{{Meixner} et~al.}{2004}]{APN3}
{Meixner} M.,  {Kastner} J.~H.,  {Balick} B.,    {Soker} N.,  eds, 2004,
  Asymetrical Planetary Nebulae III No.~313 in ASP Conf. Ser.

\bibitem[\protect\citeauthoryear{{Morisset} \& {P\'equignot}}{{Morisset} \&
  {P\'equignot}}{1996}]{MP96}
{Morisset} C.,  {P\'equignot} D.,  1996, \aap, 312, 135

\bibitem[\protect\citeauthoryear{{Morisset} \& {Stasi{\' n}ska}}{{Morisset} \&
  {Stasi{\' n}ska}}{2005}]{MS05}
{Morisset} C.,  {Stasi{\' n}ska} G.,  2005, \mnras, submitted

\bibitem[\protect\citeauthoryear{{Morisset}, {Stasi{\' n}ska} \& {Pe{\~
  n}a}}{{Morisset} et~al.}{2005}]{MSP05a}
{Morisset} C.,  {Stasi{\' n}ska} G.,    {Pe{\~ n}a} M.,  2005, \mnras, 360, 499

\bibitem[\protect\citeauthoryear{{Neiner}, {Acker}, {Gesicki} \&
  {Szczerba}}{{Neiner} et~al.}{2000}]{NAGS00}
{Neiner} C.,  {Acker} A.,  {Gesicki} K.,    {Szczerba} R.,  2000, \aap, 358,
  321

\bibitem[\protect\citeauthoryear{{O'Dell}}{{O'Dell}}{2001}]{ODell01}
{O'Dell} C.~R.,  2001, \aj, 122, 2662

\bibitem[\protect\citeauthoryear{{P{\' e}quignot}, {Ferland}, {Netzer},
  {Kallman}, {Ballantyne}, {Dumont}, {Ercolano}, {Harrington}, {Kraemer},
  {Morisset}, {Nayakshin}, {Rubin} \& {Sutherland}}{{P{\' e}quignot}
  et~al.}{2001}]{P02}
{P{\' e}quignot} D.,  {Ferland} G.,  {Netzer} H.,  {Kallman} T.,  {Ballantyne}
  D.~R.,  {Dumont} A.,  {Ercolano} B.,  {Harrington} P.,  {Kraemer} S.,
  {Morisset} C.,  {Nayakshin} S.,  {Rubin} R.~H.,    {Sutherland} R.,  2001, in
  ASP Conf. Ser. 247: Spectroscopic Challenges of Photoionized Plasmas
  {Photoionization Model Nebulae}.
p.~533

\bibitem[\protect\citeauthoryear{{Reich}, {Fuerst}, {Reich} \& {Reif}}{{Reich}
  et~al.}{1990}]{RFRR90}
{Reich} W.,  {Fuerst} E.,  {Reich} P.,    {Reif} K.,  1990, \aaps, 85, 633

\bibitem[\protect\citeauthoryear{{Vasconcelos}, {Cerqueira}, {Plana}, {Raga} \&
  {Morisset}}{{Vasconcelos} et~al.}{2005}]{2005astro.ph..6315V}
{Vasconcelos} M.~J.,  {Cerqueira} A.~H.,  {Plana} H.,  {Raga} A.~C.,
  {Morisset} C.,  2005, ArXiv Astrophysics e-prints/0506315

\bibitem[\protect\citeauthoryear{{Wen} \& {O'Dell}}{{Wen} \&
  {O'Dell}}{1995}]{WD95}
{Wen} Z.,  {O'Dell} C.~R.,  1995, \apj, 438, 784

\bibitem[\protect\citeauthoryear{{Wood}, {Mathis} \& {Ercolano}}{{Wood}
  et~al.}{2004}]{WME04}
{Wood} K.,  {Mathis} J.~S.,    {Ercolano} B.,  2004, \mnras, 348, 1337

\end{thebibliography}
\end{document}